\documentclass[onecolumn,aps,showpacs,superscriptaddress,longbibliography,nofootinbib,notitlepage]{revtex4-1}
\usepackage[normalem]{ulem}
\usepackage{color}
\usepackage{natbib}
\usepackage{braket}
\usepackage[utf8]{inputenc}
\usepackage{amsmath,amsfonts,amssymb}
\usepackage{hyperref}
\hypersetup{colorlinks,linkcolor={blue},citecolor={blue},urlcolor={red}}  

\begin{document}  
\title{Quantum and semi-quantum sealed-bid auction: Vulnerabilities and advantages}

\author{Pramod Asagodu}\thanks{pasagodu.study@gmail.com}
\affiliation{Indian Institute of Technology (Indian School of Mines) Dhanbad, Police Line Road, Hirapur, Sardar Patel Nagar, Dhanbad, Jharkhand, 826004, India}

\author{Kishore Thapliyal} \thanks{kishore.thapliyal@upol.cz}
\affiliation{Joint Laboratory of Optics of
Palack\'{y} University and Institute of Physics of CAS, Faculty of Science,
Palack\'{y} University, 17. listopadu 12, 771 46 Olomouc, Czech Republic}
\author{Anirban Pathak} \thanks{anirban.pathak@gmail.com}
\affiliation{Jaypee Institute of Information Technology, A-10, Sector-62, Noida, UP-201309, India}

\begin{abstract}
    A family of existing protocols for quantum sealed-bid  auction  is critically analyzed, and it is shown that they are vulnerable under several attacks (e.g., the participant's and non-participant's attacks as well as the collusion attack of participants) and some of the claims made in these works are not correct. We obtained the bounds on the success probability of an eavesdropper in accessing the sealed-bids. Further, realizing the role of secure sealed-bid auction in the reduction of corruption, a new protocol for sealed-bid auction is proposed which is semi-quantum in nature, where the bidders do not have quantum resources but they can perform classical operations on the quantum states. The security of the proposed protocol is established against a set of attacks, and thus it is established that the proposed protocol is free from the vulnerabilities reported here in the context of the existing  protocols.

\end{abstract}
\maketitle
\section { Introduction}
\par An auction is a process of selling goods or products to the buyer offering the highest bid (or buying from a seller offering the lowest price).  Auctioning has been relevant and known since ancient times, and its modern forms are still in use. Historically, the auction was developed as a process of negotiation enabling participants to exchange goods at the best possible value. Subsequent development and industrialization have led to different types of auction methods. The most common of these are English, Dutch, and sealed-bid auction \cite{vickrey1961auctions,mcafee1987auctions,krishna2009auction}.  In the English auction, the bidder with the highest bid wins the auction, in contrast of the Dutch auction where auction ends when either a bidder accepts the price set by the auctioneer or no participant is willing to decrease the bid anymore. 
In the English and Dutch auction, the process happens publicly so the security of the bid is not a concern. On the other hand, in the sealed-bid auction, the security of the bids is a major concern. This is so because in a sealed bid auction, all the bidders send their bids to the auctioneer simultaneously in the sealed envelopes (or equivalently in encrypted manner) so that no bidder is expected to know the bid of any other bidder until the end of the auction. Further, in most of the cases it is desirable that the bids other than the winning bid should be kept secret even after the sealed bid auction is accomplished. The relevance of auction in modern society is commemorated as the Sveriges Riksbank Prize in Economic Sciences in Memory of Alfred Nobel 2020 for ``improvements to auction theory and
inventions of new auction formats'' (see \cite{teytelboym2021discovering} and references therein for detail).

With the advent of quantum technology attempts have been made to design quantum solutions for auction.
Most of the existing works on quantum auction are focused on the sealed-bid auction. Specifically, the first quantum sealed bid auction scheme was proposed in 2009 \cite{naseri2009}. Various cryptanalysis and improvement of the Naseri's protocol \cite{naseri2009} were reported over the years \cite{qin2009cryptanalysis,yang2009improved,zheng2009comment,liu2009revisiting}.  All these protocols \cite{naseri2009,qin2009cryptanalysis,yang2009improved,zheng2009comment,liu2009revisiting} were vulnerable to the attacks from malicious bidders and a dishonest auctioneer. A post confirmation based quantum sealed bid auction was proposed as remedy \cite{zhao2010secure}. However, this scheme was also cryptanalyzed and improved further \cite{xu2011cryptanalysis,Ref38RePaper}.  These works have also motivated a set of new protocols and their improvements exploiting quantum resources for auction \cite{luo2013loophole,wang2014revisiting,zhang2018economic,sharma2017quantum}. 
In brief, various protocols for quantum sealed-bid auctions have been proposed and cryptanalyzed in the recent past. There are multiple reasons behind this thrust, firstly quantum cryptography promises to provide unconditional security (a desirable feature which is not achievable in the classical world) and the sealed-bid auction requires security whereas English and Dutch action don't require it.  Secondly, in the countries perceived to be the most corrupt according to global corruption index \cite{corruption}, many tasks are assigned or properties are sold by sealed-bid auction (specially by the Government). Consequently, a quantum secure sealed-bid auction process can potentially reduce corruption.

A secure sealed-bid auction should fulfill the following requirements \cite{shi2019privacy}:
 \begin{enumerate}
\item \textbf{Privacy:} The bid values of the bidders will not be accessible to the rest of the bidders.
\item \textbf{Verifiability:} All the bidders can verify the validity of the winning bid, i.e., a malicious bidder cannot win by changing the bid with the help of the auctioneer.
\item \textbf{Fairness:} A collusion of participants and/or the auctioneer will not allow them to win the auction.
\end{enumerate} 
Interestingly, most of the existing protocols of quantum sealed-bid auction do not satisfy the above criteria in the strict sense.  To elaborate on this point, in what follows, we will cryptanalyze a protocol of quantum sealed-bid auction proposed by Liu et al. \cite{liupaper2016} and two protocols for the same task obtained by modifying the Liu et al. protocol which were introduced by Zhang et al. \cite{zhang2018}. Specifically, Liu et al. \cite{liupaper2016} proposed a single photon based protocol of sealed-bid auction recently. Subsequently, the protocol was cryptanalyzed and improved by Zhang et al. \cite{zhang2018} who proposed two improved mechanisms of post-confirmation, which were claimed to be semi-quantum sealed-bid auction. 
The cryptanalysis performed here will establish that Liu et al. and Zhang et al. protocols are vulnerable under different attacks and thus, they do not satisfy the above listed security requirements. Further, Zhang et al.'s claim of proposing a semi-quantum sealed-bid auction is also wrong as the classical user is allowed to measure and perform operations on the superposition of classical states.

In fact, a classical user in semi-quantum cryptography is restricted to measure/prepare states in one basis set and reflect the qubits sent by the quantum user \cite{boyer2007}. Interestingly, secure semi-quantum schemes involving one or more classical users are already proposed for other cryptographic tasks, like key distribution \cite{boyer2007,boyer2009semiquantum,MSQKD}, quantum private comparison \cite{thapliyal2018orthogonal,semi2017}, online shopping \cite{semi2017}, key agreement \cite{semi2017}, two-way and controlled one-way secure direct communication \cite{semi2017}. In fact, much progress has been made in the field of semi-quantum cryptography with some experimental realization also reported recently (see \cite{iqbal2020semi} for review) which has strongly established the feasibility of the implementation of the semi-quantum schemes for various cryptographic tasks. Interestingly, Zhang et al. auction protocols \cite{zhang2018} are not semi-quantum schemes, and thus this and the existing protocols of sealed-bid auction are resource demanding; in the sense that all the users (auctioneer and bidders) are required to possess quantum resources which are costly.  These facts motivated us to propose a semi-quantum protocol of sealed-bid auction free from the vulnerabilities present in protocols of Liu et al. and Zhang et al.

The rest of the letter is organized as follows.  In Section \ref{section2}, we provide a review of Liu et al. and Zhang et al. protocols. A set of vulnerabilities of both the protocols is identified by us and those are described in Section \ref{sec:vulnerabilities}. In Section \ref{section3}, we propose a new protocol for semi-quantum sealed-bid auction. In Section \ref{section4}, we report the security analysis of the proposed protocol separately with respect to auctioneer, bidder and an external eavesdropper. Finally, in Section \ref{section5}, we conclude the work with the brief mention of a probable future vulnerability and possible remedy from that.  

\section{Review of Liu \lowercase{et al.} and Zhang \lowercase{et al.} protocols} \label{section2}

In this section, we aim to briefly describe the Liu et al. and Zhang et al. protocols before elaborating their vulnerabilities in the following section. Both the protocols have two main parts: (i)  bidding and (ii) post-confirmation parts. The Zhang et al. protocols differ from Liu et al. protocol only in the post-confirmation phase. 
Without the loss of generality, we can assume that the auctioneer is referred to as Alice, and the $N-1$ Bidders are called ${\rm Bob}_i$ ($i=1,\cdots, N-1)$. Now, we may describe Liu et al. protocol under the above considerations.

\subsection{Liu et al. protocol for quantum sealed-bid auction}

The bidding (\textbf{Steps 1-4} and \textbf{6}) and post confirmation (\textbf{Steps 5} and \textbf{7}) phases of Liu et al. protocol \cite{liupaper2016} work as follows.
\begin{itemize}
 \item[\textbf{Step 1:}] All the parties involved in the auction agree on a bid-encoding rule that a bidder would apply the Pauli operation $I\,(i\sigma_y)$ to encode 0 (1).
 \item[\textbf{Step 2:}] Alice generates $(N-1)$ number of  $m$-qubit sequences ($m$ being the bid length) $P_i:i\in{1,2,\cdots,N-1}$, where each qubit is randomly selected from $\{ \ket{0},\ket{1},\ket{+},\ket{-}\} $ with $\ket{\pm }=\frac{\ket{0}\pm\ket{1}}{\sqrt{2}}$. She randomly inserts decoy qubits  in $P_i$ to create an extended qubit sequence $P_i^\prime$ and sends each $P_i^\prime$ to the respective bidder ${\rm Bob}_i$.
 \item[\textbf{Step 3:}]  After receiving $P_i^\prime$   from Alice, ${\rm Bob}_i$ and Alice perform an eavesdropping check using the decoy qubits. If the error rate is found to be more than the acceptable limit, the protocol is discarded. They proceed if errors fewer than the tolerable rate are obtained.
\item[\textbf{Step 4:}] After dropping the decoy qubits and thus recovering the original qubit sequence $P_i$, each ${\rm Bob}_i$ encodes his bid using the rules decided in \textbf{Step 1}. Modified sequence after bid encoding is referred to as $P_i^{\prime\prime}$. 
\item[\textbf{Step 5:}] For post-confirmation, ${\rm Bob}_i$ prepares an ordered sequence of $\left\lceil\frac{m}{2}\right\rceil$ Bell states $R_{ij}: (i,j\in\{1,2,\cdots,N-1: i{\ne}j\}$) to encode the corresponding bid. The rule followed for encoding is 
\begin{equation}
    \label{EPRencoding}
    \ket{\psi^+}=00;\,\ket{\psi^-}=01;\,\ket{\phi^+}=10;\,\ket{\phi^-}=11 
\end{equation}
with $\ket{\psi^{\pm}}=\frac{\ket{01}\pm\ket{10}}{\sqrt{2}}$ and $\ket{\phi^{\pm}}=\frac{\ket{00}\pm\ket{11}}{\sqrt{2}}$.
Then ${\rm Bob}_i$ applies a permutation $\Pi_{ij}$ to his corresponding $R_{ij}$ to create $R_{ij}^\prime $ and inserts decoy qubits randomly to generate an enlarged sequence $R_{ij}^{\prime\prime}$ and sends that to ${\rm Bob}_j$. Subsequently, eavesdropping is checked using the decoy qubits and the protocol is discarded if error rate is high.
\item[\textbf{Step 6:}] ${\rm Bob}_i$ sends the encoded qubit sequence $P_i^{\prime\prime}$ back to Alice. Since she knows the initial state of the qubit sequence, she can decode the bid encoded simply by measuring in the correct basis. If the qubit state remains the same, the bid is 0, otherwise 1. After decoding all the bids, Alice announces the winner.
\item[\textbf{Step 7:}] The winner (suppose ${\rm Bob}_k$) announces his permutation operation. All the other bidders use this to recover the correct ordered sequence and use Bell measurement to decode the winner's bid. If the bid matches the announced value, the auction is considered fair, else it is declared unfair.
\end{itemize}

\subsection{Zhang et al. improved protocols for quantum auction\label{sec:Zhang}}

Zhang et al. shown that the Liu et al. protocol is vulnerable to two collusion attacks. They further improved the protocol by modifying the post-confirmation steps (i.e., \textbf{Steps 5} and \textbf{7}) in two alternate ways and that led to two improved protocols (to be referred to as Zhang et al. protocol 1 and Zhang et al. protocol 2) for quantum sealed-bid auction. The participants are further claimed to be semi-quantum. Here, we will briefly describe the Zhang et al. protocols which are improved versions of Liu et al. protocol.

\subsubsection{Protocol 1 of Zhang et al.}
In this protocol, post-confirmation steps are based on mutually unbiased basis states and the modified steps are as follows:
\begin{description}
 \item[{Step 5-Z1}] ${\rm Bob}_i$ prepares an ordered single-photon sequence $R_{ij}=\ket{r_{ij}}:\,  i,j\in\{1,2,\cdots,N-1: i \ne j\}$, and the bid is encoded such that $\ket{r_{ij}} = \ket{0}$ if the bit value is 0, else $\ket{r_{ij}} = \ket{+}$.
\item[{Step 7-Z1}]  After Alice announces the winner and the winning bid $B_{k}= \{b_{i}^k \}$, the bidders measure the $\ket{r_{ij}}$ in $Z$ basis if $b_{i}^k$ is 0 else in $X$ basis. If the measurement outcome is  $\ket{1}$ or $\ket{-}$, then the auction will be considered as unfair.\\
Previously the bidders could control their bids after post-confirmation by changing the permutation operators. However, the modified protocol does not allow the bidders this control.
 \end{description}
 
 \subsubsection{Protocol 2 of Zhang et al.}
 In this protocol, post confirmation is performed using unitary operations. Specifically, the bidders preshare an identical sequence of qubits $Q_{\rm pub} = \{\ket{q_i}, i\in \{1,2,\cdots,n\}\}$, where the sequence $Q_{\rm pub}$ is known to all the bidders that can be shared in a secure manner using keys generated by quantum key distribution.
\begin{description}
 \item[{Step 5-Z2}] ${\rm Bob}_i$ applies a unitary operation $U_{ij}(\theta_{ij})$ on the qubit sequence $Q_{\rm pub}$ before sending the encoded sequence $L_{ij}$ to ${\rm Bob}_j$. Here, the unitary 
 \begin{equation}
 \label{UnitaryOperation}
 U_{ij}(\theta_{ij})=\begin{pmatrix} \cos(\theta_{ij}) & \sin(\theta_{ij}) \\ -\sin(\theta_{ij}) & \cos(\theta_{ij}) \end{pmatrix},
 \end{equation}
where $\theta_{ij}$ depends on the bid that is to be encoded, i.e., $\theta_{ij}=\{0,\pi/4,\pi/2,3\pi/4\}$ are used to encode 00, 01, 10 and 11, respectively.  
 \item[{Step 7-Z2}] After Alice announces the winner and the winning bid, the bidders apply corresponding inverse operation U$_{ij}^{-1} (\theta_{ij})$ on $L_{ij}$ and recover the $Q_{\rm pub}^\prime$ qubit sequence. If this recovered qubit sequence $Q_{\rm pub}^\prime$ is the same as $Q_{\rm pub}$ then the auction is fair, else the auction is considered to be unfair. 
  \end{description}
 
 \section{Vulnerabilities of Liu \lowercase{et al.} and Zhang \lowercase{et al.} protocols\label{sec:vulnerabilities}}
 
 In what follows, we elaborate that Liu et al. protocol has some inconsistencies on top of the vulnerabilities already mentioned by Zhang et al. We further establish that both the modified schemes proposed by Zhang et al. cannot be performed by the semi-quantum participants. Additionally, these modified schemes fail to satisfy the fairness requirement of sealed-bid auction if the participants are not assumed semi-quantum.

\subsection{Vulnerabilities of Liu et al. protocol}

Zhang et al. pointed out the drawbacks with the post-confirmation part. Specifically, bidders use a permutation operation while post-confirmation which only the winner reveals in the end. This renders a malicious bidder and auctioneer to collude and disclose a permutation operator to pass the verifiability. Two such collusion attacks are already mentioned in \cite{zhang2018}.

Here, we mention some additional weaknesses and inconsistencies present in Liu et al. protocol which were apparently overlooked by Zhang et al. 
\begin{enumerate}
    \item A major claim of Liu et al. protocol is that in contrast to the earlier entangled-state based protocols of quantum sealed-bid auction, their protocol is single photon based, the same is emphasized on title of both the papers, but the use of Bell states in \textbf{Step 5}, makes it an inconsistent claim.

 \item In addition, in \textbf{Step 6} of Liu et al. protocol decoy qubits were not used, which allows other bidders or an external eavesdropper to alter the bid without being detected. This disturbance attack makes it difficult for the protocol to satisfy fairness as well as correctness. Interestingly, this imperfection is present in the subsequently improved Zhang et al. protocols as well.
\end{enumerate}

\subsection{Vulenrabilities of Zhang et al. protocols} 

The improvements proposed by Zhang et al. are prone to several attacks that we list here. To begin with, a semi-quantum participant \cite{boyer2007} is allowed to prepare and measure qubits in the computational basis only unlike ``they can only control single photons'' as mentioned in \cite{zhang2018}. Thus, mere modification in the post-confirmation phase of Liu et al. quantum sealed-bid auction is not expected to provide a semi-quantum scheme. Further, the modification by Zhang et al. in the post-confirmation part allows classical users to measure in more than one basis.

In what follows, we discuss the rest of the attacks on both the protocols proposed by Zhang et al. assuming that all the participants are equipped with all the quantum resources.

\subsubsection{Vulnerabilities of Zhang et al. Protocol 1}

Here, we investigate some attacks on the post-confirmation of Zhang et al. Protocol 1, which is based on mutually unbiased states. Specifically, a bidder ${\rm Bob}_j$ sends  
$\ket{0}$ and $\ket{+}$ to the rest of the bidders corresponding to the bit values 0 and 1 in the $m$-bit bid $B_{j}= \{b_{i}^j \}:\,i=1,\cdots,m$, respectively. In what follows, we would discuss all the possible attacks.

\subsubsection*{Attack 1: A semi-honest participant's attack}
A semi-honest participant is honest, but curious as he follows the protocol and attempts to gain some additional information. In this case,
a semi-honest bidder ${\rm Bob}_i$ may attempt to distinguish the non-orthogonal states $\ket{0}$ and $\ket{+}$ in $R_{ij}$ to know the bid $B_{j}$ that ${\rm Bob}_j$ has shared before deciding his own bid. {This type of state discriminating attack can be performed using two approaches, which are described below as Case A and Case B.} 

\noindent\textbf{Case A: Projective measurement of non-orthogonal states}

We know that two non-orthogonal states ($\braket{0|+}= \cos(\tfrac{\pi}{4})\neq 0$ in this case) cannot be discriminated with certainty. However, it is possible \cite{optimalPOVMbasis} to measure them in an appropriately chosen $\{\ket{\tau_{1}},\ket{\tau_{2}}\}$ basis, where 
\begin{equation}
\label{NearestOrthogonalBasis}
\begin{aligned}
	\ket{\tau_1} &= \cos( \tfrac{\pi}{8}) \ket{0} - \sin(\tfrac{\pi}{8}) \ket{1},\\
	\ket{\tau_2} &= \sin( \tfrac{\pi}{8}) \ket{0} + \cos(\tfrac{\pi}{8}) \ket{1}.
\end{aligned}
\end{equation}
${\rm Bob}_i$  may attribute measurement outcome $\ket{\tau_1}$ ($\ket{\tau_2}$) to state $\ket{0}$ ($\ket{+}$) sent by the bidder, corresponding to bit 0 (1) in the bid $B_{j}$, as
$\braket{0|\tau_1}= \braket{+|\tau_2}=\cos(\tfrac{\pi}{8})$. However, it will lead to an error with probability $p(e)=\left|\braket{0|\tau_2}\right|^2= \left|\braket{+|\tau_1}\right|^2=\frac{1}{2}(1-\sin(\tfrac{\pi}{4}))$. 

This attack allows a semi-honest participant to successfully know a bit $b_{i}^j $ with probability $p(s)=\frac{1}{2}(1+\sin(\tfrac{\pi}{4}))$. As the measurement of all the single qubit states is independent we can obtain the success rate for obtaining the whole $m$-bit bid as $[p(s)]^m$, which is significantly high unless $m$ is too large. Note that the trial to deduce each single qubit state of $R_{ij}$ can be viewed as Bernoulli distribution.

\noindent\textbf{Case B: Unambiguous state discrimination of non-orthogonal states}

A semi-honest bidder may design unambiguous discrimination of non-orthogonal states ($\ket{\psi_1}=\ket{0}$ and $\ket{\psi_2}=\ket{+}$ such that $\braket{0|+}=\cos\alpha=\cos\left(\tfrac{\pi}{4}\right)$) using the POVM \cite{optimalPOVMbasis} 
	\begin{eqnarray}
 	\label{POVMused}
 	     E_1 &= &\tfrac{1}{2\cos^2\left(\tfrac{\alpha}{2}\right)}  \ket{\psi_2^{\bot}}\bra{\psi_2^{\bot}}  = \tfrac{1}{2\cos^2\left(\tfrac{\pi}{8}\right)}  \ket{-}\bra{-},  \\  
	     E_2 &=& \tfrac{1}{2\cos^2\left(\tfrac{\alpha}{2}\right)} \ket{\psi_1^{\bot}}\bra{\psi_1^{\bot}} = \tfrac{1}{2\cos^2\left(\tfrac{\pi}{8}\right)} \ket{1}\bra{1},  \\
         E_3 &=& I-E_1-E_2.
 	\end{eqnarray}
 Notice that POVM $E_1$ is not orthogonal to $E_2$ unlike in the case of projective measurements, while $E_3$ is associated with an inconclusive outcome.
In other words, the two measurement operators ($E_1$ and $E_2$) we have chosen measure the states that have components orthogonal to the two non-orthogonal states ($\ket{\psi_2}$ and $\ket{\psi_1}$), while the third operator $E_3$ gives an inconclusive result. Therefore, the measurement outcome $E_1$ ($E_2$) implies that the qubit is in $\ket{\psi_1}$ ($\ket{\psi_2}$) state.  However, the measurement outcome  $E_3$ is obtained with probability $p_{\rm in}=\cos\left(\tfrac{\pi}{4}\right)$, and in that case  no conclusion can be made.
The value of $p_{\rm in}$ is the minimum possible value as we have used optimal POVM here \cite{optimalPOVMbasis,ivanovic1987,dieks1988,peres1988} for unambiguous state discrimination.

This attack is often discussed in the context of B92 quantum key distribution protocol \cite{bennett1992quantum} as the same encoding is used there, too (see \cite{srikara2020continuous} for discussion).

\subsubsection*{Attack 2: An outsider's or malicious participant's attack}

In this case, a malicious bidder ${\rm Bob}_i$ (Eve) may design intercept and resend attack on all the encoded strings ${\rm Bob}_j$ has sent to the rest of the bidders. He may prepare the fresh qubits in the same states he infers from the measurement and succeed in getting away with the attack undetected. On the other hand, even if he sends random qubits to all the parties auction will be terminated as unfair as the winner will fail in the post-confirmation phase.

Without loss of generality, we assume that a malicious bidder ${\rm Bob}_i$ (or equivalently Eve) intercepts $l$ of the $N-2$ copies of $m$-qubit string $R_{jx}\,: 1\leq x \leq N-1,\, x\neq j$ sent by ${\rm Bob}_j$.

\noindent\textbf{Case A: Projective measurement of non-orthogonal states}

This case of projective measurement of non-orthogonal states in $\{\ket{\tau_{1}},\ket{\tau_{2}}\}$ basis (\ref{NearestOrthogonalBasis}) to deduce each single qubit state of $R_{jx}$ from $l$ copies can be defined as
$l$ independent Bernoulli processes or binomial distribution $B(l,p(s))$. 
The malicious bidder ${\rm Bob}_i$ will assign the $k$th bit $b_{k}^j=0$ ($b_{k}^j=1$) in the recovered bid $B_{j}^{\prime}$ if corresponding $q \geq \left(\left\lfloor \frac{l}{2} \right\rfloor +1\right)$ measurements on the $k$th qubits in all $R_{jx}$ yield $\ket{\tau_1}$ ($\ket{\tau_2}$).  

This allows us to calculate the error probability in terms of cumulative distribution function of binomial distribution, i.e., $P(e)=\sum\limits_{z=0}^{\left\lfloor \frac{l}{2} \right\rfloor} \left(\begin{array}{c} l \\ z  \end{array}\right) (p(s))^z (p(e))^{1-z}$.  With increase in $l$, the probability of success $P(s)=1-P(e)$ increases considerably.  By using the properties of the binomial distribution \cite{arratia1989tutorial,ash1965information} we may obtain bounds  on the success probability as
\begin{equation}
1-\frac{1}{\sqrt{8l\frac{\left\lfloor {l}/{2} \right\rfloor}{l}\left(1-\frac{\left\lfloor {l}/{2} \right\rfloor}{l} \right)}}
\exp \left[-l D\left(\frac{\left\lfloor {l}/{2} \right\rfloor}{l}||p(s)\right)\right]\geq P(s)\geq 1-\exp \left[-l D\left(\frac{\left\lfloor {l}/{2} \right\rfloor}{l}||p(s)\right)\right],
\end{equation} 
where the relative entropy $D(x||y)=x \log \frac{x}{y}+(1-x) \log \frac{1-x}{1-y}$. 
One may verify that for $l=10$, ${\rm Bob}_i$ gets the bit $b_{k}^j$ correctly with more than 99\% probability.

Further, an independent measurement on the each qubit of $R_{jx}$ leads to the conclusion that the success rate of getting the bid $B_{j}^{\prime}=B_{j}$ is $(P(s))^m$.

\noindent\textbf{Case B: Unambiguous state discrimination of non-orthogonal states}

A malicious bidder performing unambiguous state discrimination (discussed in Case B of Attack 1) on each qubit of $R_{jx}$ from the intercepted $l$ copies can have an inconclusive outcome on all of them $P_{\rm in}=\left(p_{\rm in}\right)^l$, which becomes less than 5\% for $l=10$.
Moreover, optimal POVM results in $B_{j}^{\prime}=B_{j}$ with probability $(1-P_{\rm in})^m$, i.e., the malicious bidder ${\rm Bob}_i$ can access the whole bid information without being detected.

\subsubsection*{Attack 3: Collusion attack of participants}
In this case, $l$ bidders may collude to implement Attack 2. The mathematical details should remain the same as in Attack 2.

\subsubsection{Vulnerabilities of Zhang et al. Protocol 2}

The post-confirmation of Zhang et al. Protocol 2, which uses four states from two mutually unbiased basis sets to encode two bits of information, is also prone to some attacks mentioned hereafter. 
Specifically, a bidder ${\rm Bob}_j$ sends  
$U_{jx}(0)\ket{q_i}$, $U_{jx}\left(\frac{\pi}{4}\right)\ket{q_i}$, $U_{jx}\left(\frac{\pi}{2}\right)\ket{q_i}$, and $U_{jx}\left(\frac{3\pi}{4}\right)\ket{q_i}$ to the rest of the bidders in $L_{jx}\,:1\leq x \leq N-1,\, x\neq j$ for the 2-bits 00, 01, 10, and 11 in the $m$-bit bid $B_{j}= \{b_{i}^j \}:\,i=1,\ldots,m$, respectively. We don't think a semi-honest participant attack is possible on Zhang et al. Protocol 2. 
The use of previously shared secret $Q_{\rm pub}$ among the bidders may be useful against an external eavesdropping attack. However, it's not effective against a malicious participant or collusion attack. 
Thus, we discuss here only these attacks on Zhang et al. Protocol 2.

\subsubsection*{Attack 1: A malicious participant's attack}

Without loss of generality, we assume that a malicious bidder ${\rm Bob}_i$ intercepts $l$ of the $N-2$ copies of $m/2$-qubit string $L_{jx}\,: 1\leq x \leq N-1,\, x\neq j$ sent by ${\rm Bob}_j$. All the bidders know $Q_{\rm pub}$ which allows us to consider $\ket{q_i}=\ket{0}\,\forall i$ as the initial state. 
It results that a bidder ${\rm Bob}_j$ sends  
$\ket{0}$, $\ket{-}$, $\ket{1}$, and $\ket{+}$ in $L_{jx}$ to the rest of the bidders for the 2-bits 00, 01, 10, and 11, respectively, for sending his $m$-bit bid $B_{j}$.

${\rm Bob}_i$ measures $\frac{l}{2}$ qubits in the computational basis ($\{\ket{0},\ket{1}\}$) and the rest of the $\frac{l}{2}$ qubits in the diagonal basis  ($\{\ket{+},\ket{-}\}$). In the ideal scenario (i.e., in the absence of noise), the probability mass function of the measurement outcomes of the initial state $\ket{0}$ and $\ket{1}$ defined in the set of states $\{\ket{0},\ket{1}, \ket{+},\ket{-}\}$
will result in $\{\frac{1}{2},0,\frac{1}{4},\frac{1}{4}\}$ and $\{0,\frac{1}{2},\frac{1}{4},\frac{1}{4}\}$, respectively. Similarly, the probability distribution of the measurement outcomes for the initial state $\ket{+}$ and $\ket{-}$ defined in the set of states $\{\ket{0},\ket{1}, \ket{+},\ket{-}\}$
will result in $\{\frac{1}{4},\frac{1}{4},\frac{1}{2},0\}$ and $\{\frac{1}{4},\frac{1}{4},0,\frac{1}{2}\}$, respectively. In other words, the correct outcome is twice more likely than the rest of the results. It would be worth mentioning here that if we do not assume $\ket{q_i}=\ket{0}\,\forall i$ and in case of an arbitrary state $\ket{q_i}=\ket{\Psi}$, then ${\rm Bob}_i$ obtains the probability mass function in the set of states $\{\ket{\Psi},\ket{\Psi^{\bot}}, \ket{\Phi},\ket{\Phi^{\bot}}\}$, where $\ket{\Phi}=\tfrac{1}{\sqrt{2}}\left(\ket{\Phi}+\ket{\Phi^{\bot}}\right)$ and $\braket{\Psi|\Psi^{\bot}}=0=\braket{\Phi|\Phi^{\bot}}$.

Thus, repeating this for all the $m/2$-qubits allows ${\rm Bob}_i$ to get the bid $B_{j}$ correctly. We have already mentioned that ${\rm Bob}_i$ will not be detected in this attack.

\subsubsection*{Attack 2: Collusion attack of participants}
In this case, $l$ bidders may collude to implement Attack 2. The mathematical details should remain the same as in Attack 2.

It is worth mentioning here that Attack 2 on Zhang et al. Protocol 1 and Attack 1 on Zhang et al. Protocol 2  may be circumvented by using decoy qubit based eavesdropping checking for the secure transmission in the post-confirmation phase. However, a collusion of $l$ bidders cannot be meliorated by using decoy qubits. Further, use of decoy qubits in \textbf{Step 6} of Liu et al. protocol (and in Zhang et al. protocol as well) will also help against a disturbance attack (\cite{sharma2016verification} and references therein). In what follows, we propose a new protocol free from the stated limitations and having the advantage of the inclusion of the end-users (bidders) without quantum resources.

\section{ A semi-quantum protocol of sealed-bid auction }\label{section3}

In the introduction, we have already briefly mentioned about the semi-quantum cryptography. The semi-quantum cryptographic schemes \cite{semi2017} are exciting as they extend quantum advantage to classical users to participate in securely executing cryptographic tasks in association with quantum users. In all such schemes, some of the users are classical  in the sense that they do not have quantum resources, but they can perform classical operations on quantum states. More precisely, a classical user  is generally considered to be a participant who is capable of doing only the following operations:
\par 1. Prepare and/or measure qubits in the computational basis.
\par 2. Reflect the received qubits without disturbing the qubit states.

A classical participant performs Operation 2 as CTRL to perform eavesdropping checking, while Operation 1 is used as ENC in the protocol to be proposed here. Specifically, in what follows, in ENC, the classical user replaces the qubit sent by the quantum user with a freshly prepared qubit in the computational basis $\ket{b}$ to send message bit $b\in \{0,1\}$.  This is slightly different from the Operation 1 performed in the pioneering work of Boyer, Kenigsberg and Mor \cite{boyer2007} where semi-quantum key distribution was proposed for the first time and Operation 1 was viewed as a SHIFT operation in which a classical Bob used to measure a qubit received by him in the computational basis and subsequently prepare a fresh qubit in the same state (same as his measurement outcome) and send that to Alice. This small change does not require any additional capability and a classical Bob who can perform SHIFT operation \cite{boyer2007} can also perform ENC operation which does not require any measurement.

In the proposed protocol, the $N-1$ bidders ${\rm Bob}_i$ are considered to be classical  and the auctioneer Alice is considered to be a fully quantum  user having access to quantum hardware. 
They also share an authenticated classical channel.

Let's now describe a new protocol of semi-quantum sealed-bid auction in a few steps.
 The steps are indexed as \textbf{SPi}, where SP stands for \textit{semi-quantum protocol} and $i$th step. 
\begin{description}
\item[\textbf{SP1}]  \emph{Preparation step} - 
Before starting the main protocol for auction, classical bidders ${\rm Bob}_i$ and ${\rm Bob}_j$ share a semi-quantum key \cite{MSQKD} $K_{ij}\,\forall (i,j \in [1,N-1], i\neq j)$ mediated by Alice. In other words, Alice prepares and shares a Bell state between ${\rm Bob}_i$ and ${\rm Bob}_j$, who measure the single qubits in the computational basis randomly to obtain a shared key with the help of Alice's Bell measurement outcomes on the freshly prepared qubits sent to her in the corresponding cases. 

\item[\textbf{SP2}]  \emph{Bid commitment} - 
${\rm Bob}_i$ calculates $L_{ix}= K_{ix} \oplus H(K_{ix} \oplus B_i) \,: 1\leq x \leq N-1,\, x\neq i$, where $H(u)$ is a one-way hash function. Then he commits $L_{ix}$ to ${\rm Bob}_x$ through the authenticated classical channel between them. In the similar way, after receiving the committed bids $L_{xi}$ from all the bidders ${\rm Bob}_x$, ${\rm Bob}_i$ informs Alice. \\
${\rm Bob}_x$ further obtains $R_{ix}=H(K_{ix} \oplus B_i)$ from $L_{ix}$ using the shared key $K_{ix}$.

\item[\textbf{SP3}] \emph{Initiation} - 
On the receipt of intimation from all the bidders, Alice prepares $4m(1+\delta)$ ($\delta > 0$) qubits randomly in one of the $\{\ket{0}, \ket{1}, \ket{+}, \ket{-}\}$ states and sends them to ${\rm Bob}_i$.

\item[\textbf{SP4}] \emph{Encoding} - 
Out of all the qubit ${\rm Bob}_i$ receives from Alice, he performs ENC on $m$ qubits chosen randomly to encode $B_i$ and CTRL on the rest of the qubits. 
Subsequently, ${\rm Bob}_i$ applies a permutation operator to reorder the qubits and sends them to Alice. Notice that the classical user does not perform a measurement in this step unlike \cite{boyer2007,MSQKD,semi2017}.

\item[\textbf{SP5}]  \emph{Eavesdropping checking} - After receiving an authenticated receipt of the qubits from Alice, ${\rm Bob}_i$ announces the location and corresponding permutation operator of the CTRL qubits chosen randomly. Alice measures these qubits in the same basis as prepared by her initially. If the measurement outcome is different from the corresponding initially prepared state, then it is attributed as an error. If the error rate is more than the threshold error, the protocol is aborted. 

\item[\textbf{SP6}] \emph{Decoding the bid} - ${\rm Bob}_i$ announces the order of the ENC qubits. Alice measures ENC qubits in the computational basis and to obtain the bid. She similarly obtains all the bids and compares them before announcing the winner (say, ${\rm Bob}_k$ and the corresponding winning bid $B_k$).

\item[\textbf{SP7}] \emph{Post confirmation} - After Alice has announced the winner and the winning bid, the bidder ${\rm Bob}_x\, \forall x$ calculates the Hash Value ($R_{kx}^{\prime}=H(K_{kx} \oplus B_k)$) and compare it with the Hash Value  ($R_{kx}=H(K_{kx} \oplus B_k)$) committed in \textbf{SP2}. If $R_{kx}^{\prime} \neq R_{kx}$ for any $x$, the auction is called unfair and void, otherwise the desired task is accomplished.

\end{description}

\section{ Security analysis of the proposed  semi-quantum protocol for sealed-bid auction}\label{section4}

Any protocol of sealed-bid auction can in principle be attacked by (i) an eavesdropper Eve,  (ii)  the auctioneer Alice, and (iii) the bidder(s) Bob(s). Further, an attack may be executed by an individual or a few of  them may collude to execute the attack. In what follows, we discuss both the outsider's and participants' attacks further.

\subsection{A non-participant Eve's attacks} 

A non-participant eavesdropper would design attacks intending to get one or more bids before or after the auctioneer announces the winner. The auctioneer's broadcast of the winner is not expected to reveal information of all the bidders but the winner. Therefore, Eve has to attack the transmission of the bids from a bidder to either the auctioneer or the rest of the bidders. The communication between bidders is secured by semi-quantum key distribution \cite{MSQKD}, i.e., the hash values of the bids are encrypted by the
semi-quantum key shared between every pair of bidders. The security of a semi-quantum key distribution \cite{MSQKD} ensures the secure transmission of the hash values from an eavesdropper.

Thus, Eve can only attack a bidder to Alice transmission of the bid. One of the most relevant attacks on semi-quantum cryptography is CNOT attack. Specifically, in this attack, Eve performs a CNOT operation with target on the ancilla prepared in $\ket{0}_E$ and control on the travel qubit $\left(\alpha\ket{0}+\beta\ket{1}\right)_T\,: \left(\alpha^2 +\beta^2\right) =1$ with real parameters $\alpha, \beta \in \{0,1,\pm\frac{1}{\sqrt{2}} \}$ while Alice to the bidder transmission resulting in the state $\left(\alpha\ket{00}+\beta\ket{11}\right)_{TE}$. To pass the eavesdropping checking Eve has to perform a CNOT again while bidder to Alice transmission with control on the travel qubit and target on the ancilla resulting in $\left(\alpha\ket{0}+\beta\ket{1}\right)_{T} \ket{0}_E$. On the other hand, the message qubits in the computational basis are freshly prepared by the bidders so in this case the resulting state is $\ket{b_j}\bra{b_j}_T\left(\alpha^2\ket{0}\bra{0}+\beta\ket{1}\bra{1}\right)_{E}$, where ${b_j}\in \{0,1\}$. Thus, this attack by Eve will not be detected in the eavesdropping checking but notice that ancilla remains disentangled from the message qubits and thus Eve cannot gain any information due to this attack. However, Eve may attempt a CNOT only while bidder to Alice transmission in the same way as this results in $\ket{b_j}_T\ket{b_j}_{E}$. Eve will get all the bid information but eavesdropping checking in this case lead to detectable errors for Alice for the states prepared in the diagonal basis with probability one-half as the state after CNOT becomes $\frac{1}{\sqrt{2}}\left(\ket{00}\pm\ket{11}\right)_{TE}$. Thus, this attack is not practical for Eve as on average she will be detected with $\frac{1}{4}$ probability if Alice prepared quantum states initially in computational and diagonal basis with equal probability.

Therefore, Eve may attempt yet another attack, an intercept resend attack, to get the bids. Specifically, Eve intercepts all the qubits sent by Alice to the bidder and replaces them with his random string of qubits of the same size. She again intercepts the string when the bidder sends the encoded string back to Alice. She further resends the string intercepted from Alice back to Alice. Eve will not leave any detectable trace during eavesdropping checking by Alice and a wrong bid information will be submitted to Alice. To evade this attack by Eve the bidder may randomly measure some of the qubits sent by Alice in the computational basis to verify that he received the string sent by Alice. However, Eve may further modify her attack to make this attack more effective. Although this attack 
may be circumvented by use of a permutation operation by the bidder on the string before sending it back to Alice. The partial disclosure of the permutation operation in two steps enables legitimate bidders to disclose the bids only if errors below a threshold are obtained in the eavesdropping checking.

The permutation operation and eavesdropping checking also ensure security against disturbance attack, denial of service attack, etc. On top of that, some countermeasures are already available against usual Trojan horse attacks as well (\cite{sisodia2021optical} and references therein). When a bidder is sharing his bid with the auctioneer in a secure manner, any eavesdropping attempt may be viewed as equivalent to the outsider's attack. Therefore, privacy of the bids is ensured in the proposed scheme.

\subsection{Malicious auctioneer's attacks}

Alice may cheat in an auction either by conveying other bidders' bids to a malicious bidder or by falsely announcing a bidder as the winner. 

In the proposed protocol, Alice decodes the bids in SP6 
and thus cannot convey the other bidders' bid to a malicious bidder before he has to commit his bid in SP2.
The post-confirmation phase, i.e., SP7, used to ensure verifiability thus ensures security against any attack by a malicious auctioneer.
Therefore, Alice cannot know any bidder's bid before the malicious bidder commits his bid to the rest of the bidder.

Alice cannot announce falsely a bidder as a winner, as this cheating will be detected in the post-confirmation SP7. For instance, if the hash values committed by the winner and that of the winner's bid announced by Alice  are not equal, the auction is decided unfair.

\subsection{Participant's attack and collusion attack of participants}
  
A bidder cannot change his bid after committing it to other bidders as to change the bid, the bidder will have to find a bid higher than the maximum bid that also gives the same hash value as that committed by him. Since hash collision is not allowed, the bidder cannot find any such bid.

The bidders cannot know the bid of the other bidders except the winner announced by Alice even after the auction is accomplished successfully. A bidder only knows $L_{ij}$ but cannot extract the bid from the hash value of the bid since it is a one way function. Thus, this ensures the privacy of the loosing bidders.

Interestingly, the bidders may exploit the information they withhold regarding the permutation operation applied on the strings before sending them to Alice. Specifically, the bidders may disclose a false permutation operation intentionally to change their bids after it is sent to Alice. However, they cannot get any advantage from this attack as they had already committed their bids to the rest of the bidders before they communicate it to the auctioneer. Thus, this attack will fail in the post-confirmation phase, and thus we circumvent the attacks discussed in \cite{zhang2018} in our proposal.

Further, a collusion of more than one bidder in an auction may render the auction unfair. However,
the group of bidders colluding cannot know the bids of the rest of the bidders before committing their bids as they only know the hash values of the bids. This helps in thwarting even the collusion attacks of more than one bidder.

\section{Conclusion}\label{section5}
It is well-known that corruption can be reduced and thus development programs can be executed in a better manner if the auctions can be accomplished in a secure manner. This fact and the fact that the use of quantum resources can provide unconditional security to various quantum communication tasks (e.g., quantum key distribution \cite{bennett1992quantum,boyer2007,boyer2009semiquantum,MSQKD,srikara2020continuous}, secure direct quantum communication \cite{srikara2020continuousQSDC}, quantum dialogue \cite{banerjee2017asymmetric},  controlled secure communication \cite{thapliyal2015applications}) and secure multiparty computation tasks (e.g., quantum voting \cite{thapliyal2017protocols}, quantum e-commerce \cite{thapliyal2019quantum}) have led to  a bunch of protocols of secure sealed-bid auction \cite{luo2013loophole,wang2014revisiting,zhang2018economic,sharma2017quantum}. 
 The problem is that in the existing protocols all the bidders and the auctioneer required to have access to quantum resources which are costly. Further, there is a causal relation between corruption and GDP \cite{luvcic2016causality}, and corruption is generally more in a poorer country having lower GDP. Thus, it is not expected that all the bidders in general, and the bidders living in a low-GDP country in particular, will have the access to quantum resources. Thus, a meaningful approach to reduce corruption by using quantum resources would be to design unconditionally secure semi-quantum protocols for sealed-bid auction. Earlier an incorrect claim of designing such a protocol was made by Zhang et al. (as discussed above), but no such semi-quantum protocol satisfying all the requirements of an auction protocol was proposed. Addressing the need, in this letter, we have proposed a semi-quantum protocol for sealed-bid auction and have established that the protocol is secure against various attacks that can be performed by a malicious auctioneer and/or bidder as well as an outsider, who is neither an auctioneer nor a bidder. It is further shown that the proposed protocol is safe against collusion attacks that can be performed by the multiple legitimate users (e.g., two or more colluding bidders or the auctioneer and one or more bidders). This security analysis has further established that the proposed protocol of sealed-bid auction which requires minimal quantum resources can be used to perform a secure multiparty quantum computation task (i.e., auction) having a very strong social relevance. Further, as the protocol does not require much quantum resources, it can be implemented with the available technologies. 

In addition to the above, we have also cryptanalyzed three existing protocols of quantum sealed-bid auction, namely Liu et al. protocol and Protocols 1 and 2 of Zhang et al. The analysis has revealed that these protocols are vulnerable under various attacks and certain important claims related to the above protocols are not correct. Specifically, it is shown that neither Liu et al. protocol is a single photon based scheme nor Zhang et al. protocols are semi-quantum auction schemes. Further, Zhang et al. protocols are prone to malicious participant and eavesdropper as well as collusion of the participants. Interestingly, the proposed semi-quantum protocol is free from all these weaknesses of the previous protocols. An important observation may be that the post-confirmation phase required to attain verifiability in the semi-quantum scheme uses one-way hash functions, which accompanies the limitations of hash function, such as hash collision and possible quantum attacks on hash functions in the future. However, with the assumption that the bidders have limited quantum resources it should not affect the security of the proposed semi-quantum sealed-bid auction. Though security against an all powerful participant attack can be obtained by deciding a duration for each step of the protocol restricting the time available for the attacker. Thus, in short, the proposed protocol which is free from the limitations of the earlier protocols and secure against  the known attacks can extend the quantum advantage of unconditional security to classical bidders and  consequently reduce the corruption involved in sealed-bid auction. As the proposed scheme can be implemented with the available technologies, we conclude this letter with a hope that the experimental realization of this socially relevant quantum protocol will happen soon and that in turn will establish the usefulness of quantum technologies in the reduction of corruption involved in the auction process.

\section*{Acknowledgement}
AP acknowledges the support from the QUEST scheme of Interdisciplinary
Cyber Physical Systems (ICPS) program of the Department of Science and Technology (DST), India (Grant
No.: DST/ICPS/QuST/Theme-1/2019/14 (Q80)). KT acknowledges GA \v{C}R (project No.
18-22102S) and support from ERDF/ESF project `Nanotechnologies for Future'
(CZ.02.1.01/0.0/0.0/16\_019/0000754). 


\end{document}